# Exploring the Design Space of Lunar GNSS in Frozen Orbit Conditions

Filipe Pereira
Systems Engineering
Cornell University
Ithaca, NY, U.S.A
fmd43@cornell.edu

Daniel Selva
Aerospace Engineering
Texas A&M University
College Station, TX, U.S.A
dselva@tamu.edu

*Abstract*—We have witnessed a growing interest in lunar exploration missions in the last few years, with the announcement of NASA's Deep Space Gateway, the launch of several uncrewed missions and the involvement of private companies. However, there is currently no reliable means of accurate and instantaneous navigation in the vicinity of the Moon. This poses operational limitations on the autonomy of lunar robotic and small satellite missions. We present the preliminary results of a systems architecture study conducted on a new satellite navigation system orbiting the moon that would effectively extend the Global Navigation Satellite Systems (GNSS) space service volume to cislunar space. A constraint to lunar frozen orbit conditions under J2, C22 and third-body perturbations is added to achieve stable lunar orbits and the formulation includes the following design decisions: (1) Orbit semi-major axis, (2) Number of satellites, (3) Number of orbital planes, (4) Satellite phasing in adjacent planes, (5) Orbit eccentricity and (6) Argument of periapsis. The Borg Multi-Objective Evolutionary Algorithm (MOEA) is used to optimize the satellite constellation design using multiple crossover and mutation operators that can adaptively be selected at runtime. The fitness function takes into account performance, cost, availability and station-keeping delta-V. The performance metric assessment is based on the Geometric Dilution of Precision (GDOP), which is computed over a grid of 500 equidistant points on the lunar surface. The input satellite orbits used in the GDOP computation are obtained from high-fidelity orbit propagation using NASA's General Mission Analysis Toolbox. The space segment cost model considers satellite constellation development and production costs. Single satellite costs are based on satellite dry mass estimates derived from the power budget analysis assuming a satellite lifetime of 10 years. The results show that high-performing Lunar GNSS constellation relying on frozen orbits alone can be achieved with 20 satellites but be suboptimal for high-latitude regions.

## I. Introduction

According to the Global Exploration Roadmap [1], there is now the consensus among 14 space agencies to conduct space exploration missions in the vicinity of the moon, as a stepping-stone to the human colonization of Mars. A wide range of robotic, sample return and human lander missions to the lunar surface are planned in this decade. Moreover, important infrastructure to be deployed in the lunar vicinity such as the planned Deep Space Gateway, which is meant to act as a communications relay for smallsats, cubesats, and lunar surface assets, could transform the way those space missions are currently operated.

One major challenge in executing missions in the vicinity of the moon is accessing accurate Position, Navigation, and Timing (PNT) services. The Global Positioning System (GPS) space service volume is defined up to geostationary orbit (GEO) with the requirement that at least one satellite shall always be in view [2]. This requirement is important to ensure the permanent availability of time synchronization with GPS time. Nevertheless, there are no GPS performance guarantees above GEO and at cislunar distances. Recent studies have focused on assessing the practical limits of the current Global Navigation Satellite Systems (GNSS) space service volume. For example,[3] indicates that a typical GNSS receiver with a 20 dBHz acquisition/tracking threshold would be able to receive a GNSS signal up to half the Earth-Moon distance. This is supported by another study based on flight data from NASA's MMS mission [4]. Nevertheless, as pointed out in [5] navigation at moon distances relying on GNSS only would still need to deal with the problem of clock and range error state correlation as a consequence of poor geometry and poor visibility. Thus, Business As Usual (BAU) still relies on expensive and time-limited ground station links from Earth (e.g. Deep Space Network), which are ineffective in the case of lunar occultation.

We believe that as the number of lunar missions increases, along with the demand for high-quality Navigation services (similar to those provided by GPS on the Earth's surface) it is reasonable to consider the development of a dedicated GNSS around the moon. Such a system would be of particular interest to all missions requiring instantaneous position information anywhere on the moon's surface. We envision a system employing the same core architecture as GPS that would meet the user requirements for high-quality navigation around the Moon. By operating under the same principles, a space-qualified GNSS receiver chip could effectively be used for Positioning, Navigation and Timing services both on the Earth's and Moon's vicinity. Time synchronization is out of the scope of the present study but one reasonable solution would be to include an ensemble of atomic clocks in the Lunar Gateway to implement the Lunar GNSS system time and to synchronize it with GPS time on earth.

The problem of designing a stable constellation with good global coverage and GDOP is challenging given the combination of the weak lunar gravity field and the strong third-body perturbations from the Earth. In this preliminary study, we chose to focus on stable lunar orbits by using the analytical theory derived by Nie and Gurfil [6] for frozen orbits conditions including $J_2$, $C_{22}$, and third body perturbations. Previous studies on Lunar navigation,[7] have done detailed performance assessments not only in terms of GDOP but also taking into account dynamic solutions. However, the analysis was limited to a reduced number of architectures and without much consideration for the stability and operational feasibility of those orbits. This study attempts to explore a much wider design space and explicitly introduce station-keeping delta-v as an objective.

The rest of the paper has the following structure. Section 2 describes the methods used for architecture design and evaluation. This includes an explanation of the software setup and its interfaces, the problem formulation, objective functions, and assumptions. Finally, we describe the architecture ranking method and the indicator used to assess the GA performance. Sections 3 and 4 deal with the results and conclusions respectively.

## II. METHODS

In this study, we employ a posteriori optimization, an approach to optimization problems in which the solution search happens in advance of the decision-making process [8]. The multi-objective constellation design optimization problem presented in this paper was solved with a high-performance Multi-Objective Evolutionary Algorithm (MOEA) framework developed in ANSI C and is described in more detail below.

### A. Formulation

This study focuses on constellations of satellites characterized by a shared set of orbital elements: semi-major axis, inclination, eccentricity and argument of periapsis, that exhibit frozen orbit conditions. In particular, we focus on frozen conditions that exist when the argument of periapsis is 90 or 270 deg since these conditions are still valid for high orbit altitudes. (This is in contrast with the frozen orbit conditions found for w = 0 and 180 which are limited to an orbit semi-major axis less than 3334km). These frozen conditions impose a value of orbit inclination once the semi-major axis, eccentricity, and argument of periapsis are defined.

The exact set of individual satellite Keplerian parameters is defined once the satellites are evenly assigned to the orbital planes and the relative spacing between satellites in adjacent planes is chosen (phasing). In this regard, we use the same logic proposed by Walker [9] when describing the delta pattern.

The right ascension of the ascending nodes (RAAN) of the P distinct orbits is evenly spaced at intervals of 360°/P, where P = number of orbital planes. Additionally, the relative positions of satellites in different orbital planes are defined by an integer number F (for delta patterns it must be an integer) of pattern units (PU= 360°/T, where T is the number of satellites) called phasing. Therefore, the true anomaly is obtained by evenly spacing the satellites in the same orbital plane and adding an offset equal to F pattern units. In summary, the formulation chosen for this problem uses 6 algorithmic real variables: (1) Orbit semi-major axis, (2) Number of satellites, (3) Number of orbital planes, (4) satellite phasing in adjacent planes, (5) Orbit eccentricity and (6) Argument of periapsis. Additionally, several constraints are introduced to restrict the design space to a region of interest for GNSS constellations: (1) Number of planes must be a divisor of the number of satellites, (2) the orbits share the same argument of periapsis, (3) the satellites are evenly assigned and spaced in the orbital planes, (4) Inclination set to frozen orbit conditions for w=90,270 deg.

Orbit semi-major axis and orbit eccentricity are inherently real and treated as such in the model. The value bounds for these variables are set to [3474,17370] and [0,0.3] respectively. The algorithmic variables of the number of orbital planes $N_{PL}^{Alg}$, orbit phasing $N_{PH}^{Alg}$ and argument of perigee $N_{\omega}^{Alg}$, however, are set to a real number in the interval [0,1]. These variables are then converted to model variables to compute the objective values. The number of orbital planes is restricted to the number of distinct factors of the number of satellites ($Fact_{SV}$). To select the exact factor, $N_{PL}^{Alg}$ is scaled to the [0, $Fact_{SV}$] interval and used to pick the closest index value $\{1,.., Fact_{SV}\}$ in absolute terms. For example, if the number of satellites is 21 there are four possible options [1,3,7,21] for the number of planes, so $N_{PL}^{Alg}$ is scaled to the [0,4] interval. Assuming $N_{PL}^{Alg} = 0.6$, its scaled value would be 2.4 and the second option would be selected (i.e. $N_{PL}^{Model} = 3$). Once the number of planes $N_{PL}^{Model}$ is determined, $N_{PH}^{Model}$ is chosen using the same procedure, which results in an integer value in the range $\{0,.., N_{PL}^{Model}-1\}$. $N_{\omega}^{Model}$ can take two distinct values $\{90,270\}$, depending on whether $N_{\omega}^{Alg}$ is less or greater than 0.5 respectively.

In this study, we assume the availability of a rocket (such as SpaceX Falcon Heavy) capable of transporting the entire satellite constellation to translunar insertion orbit. The strategies for individual satellite orbit insertion are numerous and we make no assumptions in that respect. In this regard, given the presence of substantial third body perturbations there can be an interesting trade-off between deployment time and required ΔV.

### B. Metrics

The optimization problem as formulated above is solved by the genetic algorithm, which attempts to find the set of optimal solutions according to a predefined set of objectives. In this study the goal is to (1) minimize the Geometric Dilution of Precision (GDOP), (2) Maximize system Availability on the lunar surface (3) Minimize the total space segment development and production costs, (4) Minimize the required orbit station-keeping ΔV.

#### 1) Geometric Dilution of Precision

We consider a Lunar GNSS relying on 1-way pseudorange and carrier phase measurements, similarly to the way GPS works, but we make no assumptions in terms of the User Equivalent Range Error (UERE). Due to the absence of atmosphere on the Moon, UERE would likely be driven by the satellite clock and ephemerides error, which depend on the ground and control segment designs -out of the scope of the current analysis. Nevertheless, a good estimation of the User Navigation Error (UNE) under cold start conditions (i.e., no a priori knowledge of one's position) can be achieved by multiplying the GDOP with a sensible value of UERE. Thus, the GDOP - indication of satellite geometric diversity - represents the main performance metric used in this study.

The GDOP is computed at 5 min intervals over 500 possible user locations on a spherical grid of equidistant points with a radius corresponding to the mean lunar equatorial radius (Fig. 1). The 98th percentile of GDOP values less than 6.0 obtained during a 24h period over all locations is used as the objective value. We chose a 24h period starting at 1 sidereal month after the initial orbit definition epoch. This is because the GDOP value obtained was found to be within 1% of the GDOP obtained when considering the whole first month period but lead to computation time savings of ~40%.

*2) Availability*

One of the GNSS key performance indicators is Availability, which provides an assessment of how usable the system is according to some definition of usability. In this study, we assume that the system is usable at a specific user location if the GDOP is less than 6.0. Thus, we compute the Availability of service as the percentage of the time this condition is verified across all the 500 different user locations on the lunar surface.

*3) Space segment costs*

The costs associated with the deployment of a Lunar GNSS constellation are numerous and difficult to estimate without numerous assumptions. The objective in this study is to have an architecture distinguishing metric that captures the order of magnitude of the space segment cost. Thus, we estimate the development cost, $Cost_{dev}$, and production cost, $Cost_{prod}$, of the entire satellite constellation using first principles and empirical functions derived from previous space missions.

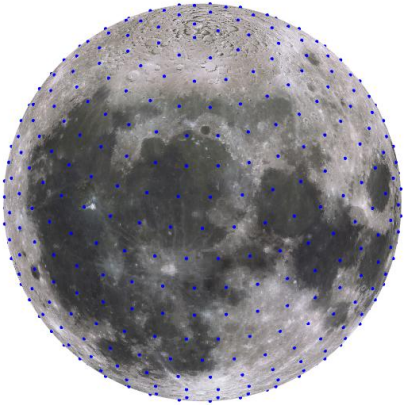

Fig. 1. Simulated user locations on the lunar surface (for GDOP computation)

$$Cost_{space\ segment} = Cost_{dev} + Cost_{prod} \quad (1)$$

Assuming a learning curve (S) of 85% and a satellite constellation with N satellites the production costs ($Cost_{prod}$) are given by:

$$Cost_{prod} = T_1 N^{(1+\log S/\log 2)} \quad (2)$$

where $T_1$ is the first unit production cost, which is computed with the USCM8 Cost Estimating Relationship (CER) [10] as a function of the satellite dry mass $m_{dry}$:

$$T_1 = (289.5 m_{dry}^{0.716})/1000 \quad (3)$$

The costs of development plus on qualification unit ($Cost_{dev}$) are also given by USCM8 as:

$$Cost_{dev} = (110.2 m_{dry})/1000 \quad (4)$$

Both cost estimates are given in FY2010 in millions of US dollars. The SEE of (3) inside the domain 288-7398kg is 21%. The standard error of the estimate (SEE) of (4) inside the domain 114 – 5127kg is 47% [10].

To have an accurate estimation of satellite dry mass we use an empirical function that depends on payload power consumption ($P_{PL}$) as well as on propellant mass. The propellant mass is obtained with the rocket equation taking the estimated station-keeping ΔV over 10 years as input and assuming the use of hydrazine monopropellant with a specific impulse $I_{sp}$= 227s.

$$m_{dry} = 38(0.14 \cdot P_{PL} + m_{propellant})^{0.51} - m_{propellant} \quad (5)$$

$$m_{propellant} = m_{dry}^{init}\left(\exp\left(\frac{\Delta V_{10\ years}}{9.8 I_{sp}}\right) - 1\right) \quad (6)$$

$$m_{dry}^{init} = 7.5 \cdot P_{payload}^{0.65} \quad (7)$$

Equations 5 and 7 are derived in [11] from empirical models that take into account data from past communication satellite missions between 1990 and 1999. To have a rough initial estimate of satellite dry mass for orbit propagation (GMAT input) and propellant mass estimation (7) is used.

The satellite payload power is computed assuming a payload architecture inspired in the GALILEO Full Operational Capability (FOC) satellite and consisting of the following components : (1) two Rubidium Atomic Frequency Standard (RAFS) atomic clocks,(2) two Phase Hydrogen Maser (PHM) atomic clocks, (3) Travel wave tube amplification with 68% efficiency, (4) Frequency Generation and Upconversion Unit (FGUU), (5) Navigation Signal Generation Unit (NSGU), and (6) Remote Terminal Unit (RTU). The corresponding power consumption of each component is summarized in Table 1. Furthermore, it is assumed that the P/L thermal subsystem, $P_{Th}$, requires 15% of the estimated value. The total payload power estimate, $P_{PL}$, is obtained by taking the maximum power consumption values for each of the main components, as shown in (8). The actual payload consumption may be lower, but this conservative estimate allows us an additional margin for unquantified losses.

$$P_{PL}[W] = \frac{P_T}{0.68} + P_{PHM} + P_{RAFS} + P_{FGUU} + P_{NSGU} + P_{RTU} + P_{Th} \quad (8)$$

$P_T$ is the satellite transmit power necessary to achieve a minimum received signal power of -150dB on the lunar surface as determined by the link budget in (9). The following assumptions are made: (1) the satellite transmits a single frequency signal at L1 ($f$=1575.42 MHz) through an L-band antenna with a gain, $G_T$, of 13 dBi, (2) antenna polarization loss $L_{ant}$ of 2 dB, (3) excess loss $L_{ex}$ of 0.5 dB and (4) user antenna gain, $G_R$, of 0 dBi:

$$P_T = P_R - G_T - G_R + L_{ex} + L_{ant} - 20\log_{10}(c/(4\pi \cdot f \cdot r_{s2r}^{max})) \quad (9)$$

The maximum user range, $r_{s2r}^{max}$, is computed for the case of a satellite transmitting the navigation signal at apoapsis ($a_{apo}$) and a user receiving the signal at the horizon (elevation angle $\eta = 0°$). These correspond to the worst-case conditions.

$$r_{s2r}^{max} = -r_m \sin(\eta) + \sqrt{a_{apo}^2 - r_m^2 \cos^2(\eta)} \quad (10)$$

where $r_m = 1738.1\ km$ is the mean moon radius.

*4) Station-keeping Delta-V*

One of the main challenges of designing a satellite constellation in lunar orbit is the need to rigorously assess the influence of third body perturbations. The influence of the Earth's gravitational potential is very significant and can result in unstable orbits especially as the orbit altitude increases. Furthermore, the maintenance of satellite geometry is essential for navigation performance guarantees and it makes sense that the constellation design takes into account the ΔV required for station keeping maneuvers throughout the satellite's lifetime. Therefore, in this study, one of the objectives that we are trying to minimize is the required station keeping ΔV required to keep the satellite eccentricity within 0.8%, the argument of periapsis within 1 deg (in the case of eccentricity > 0.1) and the magnitude of the apoapsis position vector to within 1 km of the specified goal. Whenever the eccentricity or argument of periapsis fall out of bounds, GMAT automatically computes the ΔV necessary to execute two instantaneous maneuvers and correct the trajectory to the desired orbit. GMAT is configured to use the Newton Raphson algorithm for solving this boundary value problem.

TABLE I. PAYLOAD POWER CONSUMPTION

| Payload component | Units [#] | Maximum power consumption [W] | Ref. |
|---|---|---|---|
| PHM, $P_{PHM}$ | 2 | 54 | [12] |
| RAFS, $P_{RAFS}$ | 2 | 39 | [13] |
| NSGU, $P_{NSGU}$ | 1 | 35 | [14] |
| FGUU, $P_{FGUU}$ | 1 | 22 | [15] |
| RTU, $P_{RTU}$ | 1 | 12 | [16] |

*C. Software setup*

The fitness function module containing the objective definitions was developed in MATLAB and the MOEA interface is executed through a MATLAB wrapper. Furthermore, the orbit propagation is performed using the General Mission Analysis Toolbox (GMAT), NASA's open-source high fidelity orbit propagator. The interface is made through the use of scripts also written in MATLAB as depicted in Fig 2. This software setup was run in an 8-core laptop computer with a 2.3 GHz processor and 32 GB RAM. To reduce the execution time of each function evaluation, the most computationally expensive parts of the code (GDOP computation and Orbit propagation) were parallelized to take advantage of the 8 available cores. Nevertheless, each function evaluation took approximately five minutes to execute.

*1) Borg MOEA*

The multi-objective optimization problem presented in this paper is studied with the Borg (MOEA) Framework [17]. This framework includes an auto-adaptive crossover operator selection that is particularly versed to explore the objective space in multimodal problems. Borg MOEA is a top performer in many benchmark problems (e.g, UF2, DTLZ2, DTLZ4) and it contains several important features such as ε-dominance, polynomial mutation, ε -dominance archive and ε-progress (a measure of convergence speed that triggers restarts if needed). The following set of crossover operators is available: Simulated Binary crossover (SBX), Differential Evolution (DE), Parent-Centric crossover (PCX), Unimodal Normal Distribution crossover (UNDX), Simplex crossover (SPX) and Uniform Mutation (UM). One important feature of Borg is the inclusion of a restart mechanism for reviving search once stagnation is detected with ε-progress. This involves (1) the adaptation of the search population size to remain proportional to the archive size, (2) the adaptation of the tournament selection size to maintain elitist selection, and (3) the population is emptied and repopulated with solutions from the archive.

The epsilon values are important to determine the search resolution in EA since it determines the population that it is selected to be included in the archive, which in turn is used to produce the population in subsequent generations. The epsilon values chosen for the four objective variables were 0.01, 0.1, 10 and 0.01 for GDOP, Availability, Cost, and ΔV.

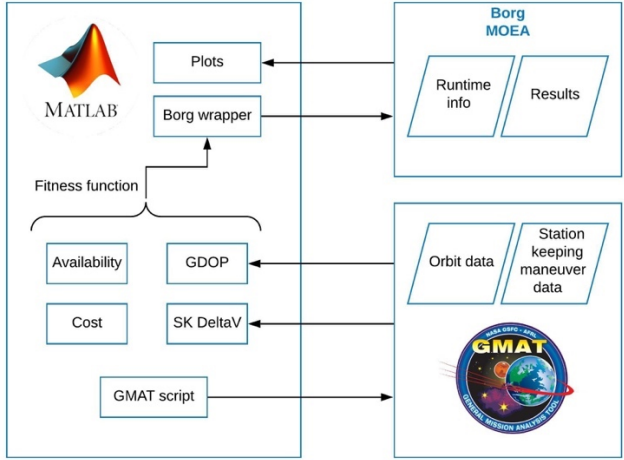

Fig. 2. Software setup

*2) GMAT*

The satellite orbits used in the GDOP computation are obtained from high-fidelity orbit propagation using NASA's General Mission Analysis Toolbox (GMAT). GMAT is configured to run a variable-step PrinceDormand78 propagator with the LP165P lunar spherical harmonic model (configured with degree and order 10) and the DE405 planetary ephemerides. Third body (Earth and Sun) and solar radiation pressure perturbations are considered assuming a satellite reflectivity coefficient of 1.8 and a cross-sectional area of $3m^2$. The satellite dry mass value used in the orbit propagation runs is a function of payload power and propellant mass as described in more detail below. The GMAT mission sequence is configured to activate two independent instantaneous maneuvers to maintain the target orbit parameters with some degree of tolerance. These maneuvers happen when one of two conditions take place: (1) the absolute value of the orbit eccentricity exceeds 0.8% w.r.t the original value (2) the absolute value of the argument of periapsis exceeds 1 degree with respect to the original value. The maneuvers if activated will attempt to bring the orbit eccentricity, argument of periapsis and magnitude of the position vector at apoapsis to the specified values at the initial time epoch.

*D. Pareto sorting*

After completion of the MOEA runs the solutions are sorted by Pareto ranking. The sorting algorithm finds the Pareto front, which is composed of all the solutions that are better than all alternatives in at least one objective while being no worse in the other objectives. The resultant Pareto optimal solutions are hence the set of non-dominated solutions and assigned a rank of one. The process continues by iteratively finding the nondominated front among the unranked solutions and incrementing the rank by one.

*E. Hypervolume indicator*

The Hypervolume indicator measures the hypervolume of objective space dominated by a non-dominated set, thus representing a combination of both convergence and diversity. This indicator has the advantage of being intuitive, compatible with outperformance relations and scale-independent. [18]

## III. RESULTS

In this section, we present the results obtained after running the MOEA for approximately 3 weeks. The Pareto front obtained after almost 4000 function evaluations out of two different seeds totaled 514 architectures, which are shown with a filled circle in Fig. 5. Despite the limited number of functions evaluations, the algorithm kept converging and did not deteriorate —The fact that the hypervolume indicator shown in Fig. 3 is monotonically increasing means that the algorithm is not losing non-dominated solutions over time.

Some interesting insights can be obtained by looking at the Pareto front in Fig 4. The solutions are color-coded by station-keeping ΔV and it is possible to see that a large number of architectures require less than 0.2km/s per satellite per year. The mean and standard deviation of ΔV values is 0.41 and 0.70 km/s per satellite per year respectively. For comparison, the current GPS satellites require ~0.12m/s/ year.

The fact that there are architectures with much larger ΔV despite all architectures obeying the same frozen orbit conditions is expected since the high-fidelity dynamics in the simulation are not matched in the derivation of the close-form analytical solutions. Notably, the frozen orbit model assumes that the Earth is in a circular equatorial orbit in the Lunar inertial frame[6].

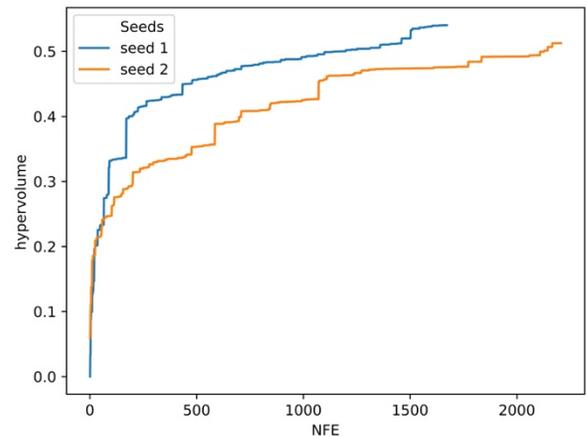

Fig. 3. Hypervolume indicator vs Number of function evaluations

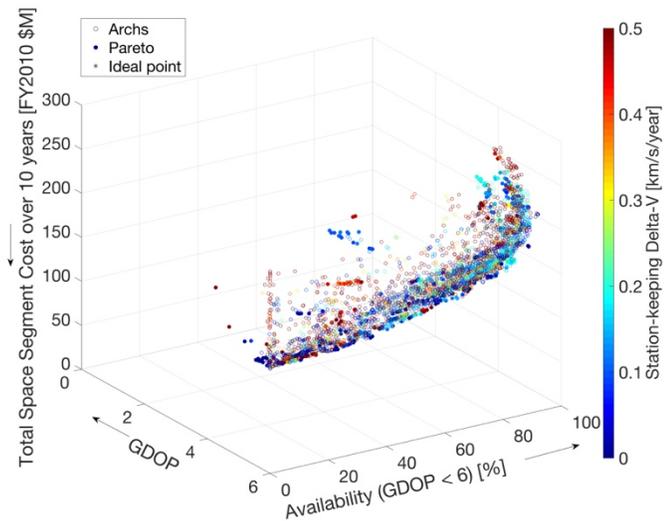

Fig. 4. Tested architectures in objective space for all 4 objectives (colored by station-keeping ΔV)

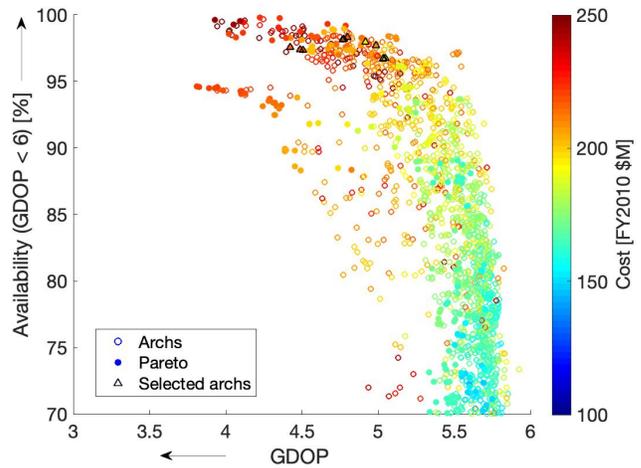

Fig. 5. Availability vs GDOP (colored by space segment cost)

A few more insights can be derived from the 2D plot in Fig 5. which color-codes the solutions by space segment costs while ignoring the station-keeping ΔV objective. As expected, the space segment costs increase as we approach the availability requirement for global coverage, i.e., 100%, since the size of the constellation also increases. Based on the results, a minimum of 18 satellites is needed for 90% availability vs 20 satellites for 95%. It also visible that the best performing constellations in terms of GDOP do not achieve more than 95% availability. This could be because we haven't explored the entire objective space. If not, the decision-maker may prefer a better performing constellation if the lack of availability occurs in relatively uninteresting regions on the lunar surface. Highlighted in Fig.5 are several Pareto front architectures that we consider more interesting since they ensure at least 95% availability, a station-keeping ΔV of less than 0.1 km/s per satellite per year and are characterized by less than 10 orbital planes. The preference for constellations with less than 10 orbit planes is grounded in orbit insertion considerations, since the lower the number of planes the easier it is to deploy the constellation.

In Table I, we highlight two constellations that constitute a good trade-off between constellation size, number of planes and the objectives. Architecture ID 1 exhibits the lowest number of satellites (20) in 5 planes at roughly 8000km semi-major axis, while architecture 7 requires 21 satellites in 3 orbital planes at roughly 7500km semi-major axis. These architectures also contain very small station-keeping ΔV and eccentricity values. Most of the architectures in Table I are nearly circular, which constitutes an additional advantage for a satellite navigation system since the received power on the lunar surface would be essentially constant.

One consequence of the frozen orbit condition is that the orbit inclination for all the optimal architectures in Table I is set at around 40 deg. This restriction imposes GDOP degradation at higher latitudes, especially at the poles as can be seen in Fig. 6. As a result, it would be interesting to consider e.g., a hybrid constellation composed of these frozen orbits in addition to a few satellites in polar orbit.

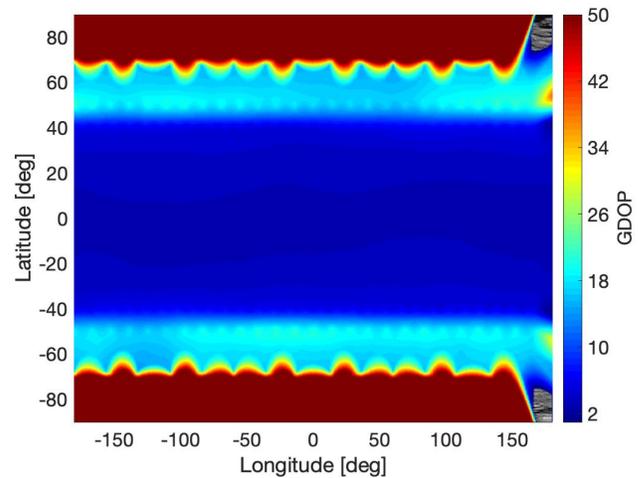

Fig. 6. GDOP (98th percentile) on the lunar surface over one month for architecture ID 1

## IV. CONCLUSIONS

### A. Contributions

The analysis reported here is, as far as we know, the first attempt to explore the Lunar GNSS design space under frozen orbit conditions and find the optimal architectures in terms of performance (GDOP), availability (percentage of time with GDOP < 6.0), space segment cost and orbit stability (by means of a low station-keeping ΔV requirement). Such a Lunar GNSS would be most valuable as a primary means of navigation for missions requiring instantaneous PNT services anywhere on the lunar surface.

Our preliminary assessment sheds new light on the potential of frozen orbits to integrate the design of a Lunar GNSS. First, the analysis indicates that it is possible to design a good GNSS constellation of satellites in frozen orbits alone. For instance, a constellation of 20 satellites in near-circular orbit at ~6200km altitude (semi-major axis ~ 8000km) can produce a GDOP (98% percentile) of 5.0, an Availability of ~97% and require a station-keeping ΔV of 0.07 km/s per satellite per year at a cost of ~200 $M. Second, we show that there is a significant variation ($\mu = 0.41$, $\sigma = 0.70$) in the values of station-keeping ΔV, even though that all architectures are subject to the same frozen orbit constraints. This is likely due to the varying sensitivities of the satellite orbits to the eccentricity and inclination of the Earth's orbit —The Earth is the largest perturbation at an altitude higher than 1500km [6]—, which is modeled in the simulation but ignored in the derivation of the analytical equations. Third, the frozen orbit conditions impose a limitation on the orbit inclinations of around 40 deg at the altitude regimes that exhibit the best navigation performance. This is detrimental to the achieved performance at higher latitudes, especially at the poles.

### B. Limitations

Despite the interesting preliminary results presented in this paper, there are numerous limitations. First, we could only afford a limited number of function evaluations, and seeds which did not allow the exploration of the full Pareto front. Second, the analysis relies on frozen orbits alone, which may not lead to the best performing and cost-effective architectures. Third, we did not take into account the operational costs related to the constellation deployment in particular with orbit insertion. Finally, we don't know how robust the optimal constellations are under different assumptions (e.g., minimum received power) which could be achieved with an analysis of sensitivity over key parameters.

### C. Future Work

We plan to remove the frozen orbit constraint and change the formulation to enable a more generic description of the problem. This formulation would allow the consideration of hybrid architectures, which may outperform the architectures presented in this preliminary study (e.g., a constellation with a subset of satellites in frozen orbits and a subset in polar orbit). To effectively explore the Pareto front in this demanding scenario, we plan to convert the MATLAB code to C++ for efficiency and use a computer cluster to significantly lower the required time per function evaluation.

TABLE II. SELECTED ARCHITECTURES

| | Architectures | | | | | | | Objectives | | | | Additional Information | | | |
|---|---|---|---|---|---|---|---|---|---|---|---|---|---|---|---|
| ID | SMA [km] | # SV | # Planes | Phasing | Eccentricity | Inclination [deg] | Argument of periapsis | GDOP | Avail [%] | Cost [$M] | ΔV [km/s /year] | SV dry mass [kg] | Transmit Power [dB] | Payload power [W] | Flight unit cost [FY10$M] |
| 1 | 8025.9 | 20 | 5 | 0 | 0.004 | 39.53 | 270 | 5.01 | 97.06 | 189.47 | 0.07 | 273.17 | 13.78 | 341.63 | 16.07 |
| 2 | 8148.8 | 20 | 5 | 0 | 0.004 | 39.51 | 90 | 5.03 | 96.86 | 189.80 | 0.01 | 273.72 | 13.92 | 342.98 | 16.10 |
| 3 | 7298.6 | 21 | 3 | 1 | 0.001 | 39.71 | 270 | 5.26 | 95.83 | 193.69 | 0 | 269.98 | 12.91 | 333.86 | 15.94 |
| 4 | 8669.2 | 24 | 4 | 0 | 0.024 | 39.46 | 270 | 4.49 | 97.38 | 215.47 | 0.07 | 276.98 | 14.48 | 350.99 | 16.23 |
| 5 | 8916.6 | 24 | 4 | 1 | 0 | 39.41 | 90 | 4.78 | 98.31 | 215.76 | 0.05 | 277.14 | 14.73 | 351.43 | 16.24 |
| 6 | 8904.4 | 24 | 4 | 1 | 0 | 39.41 | 90 | 4.77 | 98.12 | 215.53 | 0.03 | 277.09 | 14.72 | 351.3 | 16.24 |
| 7 | 7434.8 | 21 | 3 | 1 | 0 | 39.67 | 270 | 5.29 | 95.87 | 193.98 | 0 | 270.51 | 13.08 | 335.15 | 15.96 |
| 8 | 7298.6 | 21 | 3 | 1 | 0.001 | 39.71 | 90 | 5.25 | 95.80 | 193.69 | 0 | 269.97 | 12.91 | 333.85 | 15.94 |
| 9 | 8954.2 | 24 | 4 | 1 | 0.002 | 39.40 | 90 | 4.80 | 98.31 | 215.73 | 0.05 | 277.43 | 14.77 | 352.14 | 16.25 |
| 10 | 8536.0 | 24 | 4 | 0 | 0.025 | 39.47 | 270 | 4.51 | 97.36 | 215.10 | 0.07 | 276.35 | 14.34 | 349.44 | 16.21 |
| 11 | 5701.2 | 24 | 4 | 1 | 0.002 | 40.78 | 90 | 4.99 | 97.71 | 207.88 | 0 | 264.16 | 10.59 | 319.9 | 15.69 |
| 12 | 8855.4 | 24 | 4 | 0 | 0.023 | 39.43 | 270 | 4.42 | 97.55 | 215.98 | 0.08 | 277.85 | 14.67 | 353.17 | 16.27 |
| 13 | 8904.4 | 24 | 4 | 1 | 0 | 39.41 | 90 | 4.77 | 98.12 | 215.53 | 0.03 | 277.09 | 14.72 | 351.3 | 16.24 |